\documentclass[runningheads]{svmult}

\usepackage{makeidx}   
\usepackage{graphicx}  
\usepackage{subeqnar}  
\usepackage{multicol}  
\usepackage{physprbb}  
\makeindex             

\def\Journal#1#2#3#4{{#1} {\bf #2}, #3 (#4)}

\def\PRC{{\em Phys. Rev.} C}

\def\A&A{\em Astron. and Astrophys.}

\def\NPA{{\em Nucl. Phys.} A}

\def\ApJ{{\em Astrophys. Journ.}}
\newcommand{\beq}{\begin{equation}}
\newcommand{\eeq}{\end{equation}}
\def\bar{\begin{array}}
\def\ear{\end{array}}
\def\cap{\par\noindent}
\def\bra{\langle}
\def\ket{\rangle}
\def\le#1{\label{eq:#1}}
\def\re#1{\ref{eq:#1}}

\begin{document}
\title*{Microscopic Theory of the Nuclear Equation 
 \protect\newline of State and Neutron Star Structure}
\toctitle{Microscopic Theory of the Nuclear Equation 
 \protect\newline of State and Neutron Star Structure}
%
%
\titlerunning{Nuclear Equation of State}
%
\author{Marcello Baldo       
\and Fiorella Burgio}         
\authorrunning{M. Baldo and F. Burgio}
%
%
\institute{Istituto Nazionale di Fisica Nucleare, Sez. Catania, 
and Universit\'a di Catania, \\
     Corso Italia 57,
     95129 Catania, Italy}

\maketitle              

\begin{abstract}
The Bethe-Brueckner-Goldstone many-body theory of the Nuclear Equation of 
State is reviewed in some details. In the theory, one performs an expansion 
in terms of the Brueckner two-body scattering matrix and an ordering of the 
corresponding many-body diagrams according to the number of their hole-lines. 
Recent results are reported, both for symmetric and for pure neutron matter, 
based on realistic two-nucleon interactions. It is shown that there is
strong evidence of convergence in the expansion. Once three-body forces are 
introduced, the phenomenological saturation point is reproduced and the theory
is applied to the study of neutron star properties. One finds that in the 
interior of neutron stars the onset of hyperons strongly softens the
Nuclear Equation of State. As a consequence, the maximum mass of neutron 
stars turns out to be at the lower limit of the present phenomenological 
observation.
\end{abstract}

\section{Introduction}
It is believed that macroscopic portions of (asymmetric) nuclear matter 
form the interior bulk part of neutron stars, commonly associated with
pulsars. Despite infinite nuclear matter is obviously an idealized physical 
system, the theoretical determination of the corresponding Equation
of State is, therefore, an essential step towards the understanding
of the physical properties of neutron stars. On the other hand,
the comparison of the theoretical predictions on neutron stars with the
experimental observations can provide serious constraints on the
Nuclear Equation of State.    
 Unfortunately, neutron stars are elusive astrophysical
objects, and only indirect observations of their structure, 
including their sizes and masses, are possible. However, the
astrophysics of neutron stars is rapidly developing, in view of 
the observations coming from the new generation of artificial satellites, 
and one can expect that it will be possible in the near future
to confront the theoretical predictions with more and more stringent
phenomenological data.\par
Heavy ion reactions is another field of research where the nuclear
Equation of State (EOS) is a relevant issue. In this case, the
difficulty of extracting the EOS is due to the complexity of the
processes, since the interpretation of the data is necessarily 
linked to the analysis of the reaction mechanism. An enormous 
amount of work has been done in the last two decades in the field, 
but clear indications about the main characteristics of the EOS have 
still to come. Furthermore, the typical time scale of heavy ion reactions
is enormously different from the typical neutron star time scale,
and this can prevent a direct link between the two field of research.
In particular, nuclear matter inside neutron stars is completely
catalized, namely it is quite close
to the ground state, reachable also by weak processes. 
In heavy ion reactions the evolution
is too rapid to allow weak processes to relax the system towards
such a catalized state, and therefore the tested Equation
of State can differ from the neutron star one, especially at
high density. \par
On the theoretical side, the main general difficulty is the treatment of the 
strong repulsive core, which dominates the short range behaviour of 
the nucleon-nucleon (NN) interaction, typical of the nuclear system, but
which is common to other systems like liquid helium. 
Simple perturbation theory cannot of course be applied,
since the matrix elements of the interaction are too large.
One way of overcoming this difficulty is to introduce the two-body
scattering G-matrix, which has a much smoother behaviour even for
strong repulsive core. It is possible to rearrange the perturbation
expansion in terms of the reaction G-matrix, in place of the original
bare NN interaction, and this procedure is systematically exploited
in the Bethe-Brueckner-Goldstone (BBG) expansion \cite{book}.
In this contribution we present the 
latest results on the nuclear EOS based on BBG expansion and their 
applications to the physics of neutron stars.

\section{The BBG expansion and the nuclear EOS}

The BBG expansion for the ground state energy at a given density, 
i.e. the EOS at zero 
temperature, can be ordered according to the number of independent
hole-lines appearing in the diagrams representing the different
terms of the expansion. This grouping of diagrams generates the
so-called hole-line expansion \cite{Day}. The smallness parameter
of the expansion is assumed to be the ``wound parameter'' \cite{Day}, 
roughly determined by the ratio between the core volume and the volume per
particle in the system. It gives an estimate of the decreasing factor 
introduced by an additional hole-line in the diagram series.   
The parameter turns out to be small enough up to
2-3 times nuclear matter saturation density. The diagrams with a given
number $n$ of hole-lines are assumed to describe the main 
contribution to the $n$-particle correlations
in the system. At the two
hole-line level of approximation the corresponding summation
of diagrams produces the Brueckner-Hartree-Fock (BHF) approximation,
which incorporates the two particle correlations.
The BHF approximation includes the self-consistent
procedure of determining the single particle auxiliary potential,
which is an essential ingredient of the method.
Once the auxiliary self-consistent potential is introduced,
the expansion is implemented by introducing the set of diagrams
which include ``potential insertions". To be specific, the introduction 
of the auxiliary potential can be formally performed by splitting the
hamiltonian in a modified way from the usual one
\beq
H = T + V = T + U + (V - U) \equiv H_0' + V'
\le{split}
\eeq 
\noindent
where $T$ is the kinetic energy and $V$ the nucleon-nucleon 
interaction. Then one consider $V' = V - U$ as the new interaction potential
and $H_0'$ as the new single particle hamiltonian. Then,
the single particle energy $e(k)$ is given by
\beq
e(k) = {\hbar^2 k^2\over 2m} + U(k)
\le{spe}
\eeq
\cap
while $U$ must be chosen in such a way that the new interaction $V'$
is, in some sense, ``reduced" with respect to the original one $V$, so that
the expansion in $V'$ should be faster converging. The introduction of the
auxiliary potential turns out to be essential, otherwise the 
hole-expansion would be badly diverging. The total energy $E$ can then
be written as 
\beq
 E = \sum_k e(k) + B
\le{ene}
\eeq
\noindent
where $B$ is the interaction energy due to $V'$. 
\par
\begin{figure}[ht]
\vspace{-3.4 cm}
\begin{center}
\includegraphics[width=1.3\textwidth]{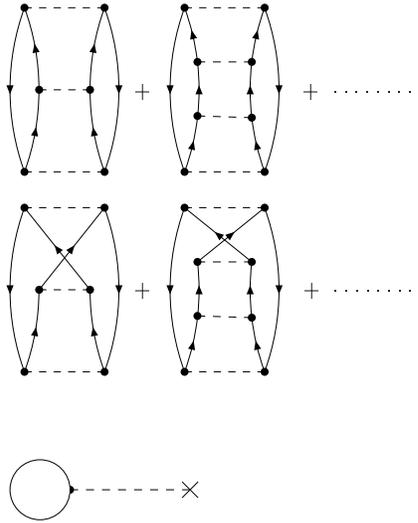}
\end{center}
\vspace{-11.2 cm}
\caption[]{Third and forth order ladder diagrams in the bare
interaction (dashed lines) and first order potential 
  insertion (bottom).}
\label{ladder}
\end{figure}

 
The BHF sums the so called
``ladder diagrams". Some of them are depicted in Fig.~\ref{ladder}.
One has to consider this set of diagrams where one, two, three, and
so one, two-body interactions $v$ appear, including exchange terms.
Special care must be used in counting correctly the diagrams which
give the same contribution.

\cap
The repeated action of the two-body
potential $v$ clearly describes the scattering of two nucleons
which lie above the Fermi sphere.  
The summation of the ladder diagrams can be performed by 
solving the integral equation for the Brueckner G-matrix
\beq
\bar{rl}
 \bra k_1 k_2 \vert G(\omega) \vert k_3 k_4 \ket\!\!\! &\, = \, 
 \bra k_1 k_2 \vert v \vert k_3 k_4 \ket \, +\,  \\
 &                  \\
 + \sum_{k'_3 k'_4} \bra k_1 k_2 \vert v \vert k'_3 k'_4 \ket\!\!\! 
 &{\left(1 - \Theta_F(k'_3)\right) \left(1 - \Theta_F(k'_4)\right)
  \over \omega - e_{k'_3} - e_{k'_4} }
  \, \bra k'_3 k'_4 \vert G(\omega) \vert k_3 k_4 \ket \ \ 
\ear
\le{bruin}
\eeq
\cap
where $\Theta_F(k) = 1$ for $k < k_F$ and is zero otherwise, being
$k_F$ the Fermi momentum. 
The product $Q(k,k') = (1 - \Theta_F(k)) (1 - \Theta_F(k'))$,
appearing in the kernel of  Eq. (\re{bruin}), enforces the scattered
momenta to lie outside the Fermi sphere, and it is commonly referred
as the ``Pauli operator". This G-matrix can be viewed as the in-medium
scattering matrix between two nucleons. It has to be stressed that
the scattering G-matrix depends parametrically on the entry energy
$\omega$, namely it is defined in general also off-shell, as the usual
scattering matrix in vacuum.
The self-consistent single particle potential 
$U(k)$ is determined by the equation
\beq
 U(k) = \sum_{k' < k_F}
      \bra k k' \vert G(e_{k_1}+e_{k_2}) \vert k  k' \ket_A \ \ \ 
\le{auxu}
\eeq
\cap
with $\vert k k' \ket_A = \vert k k' \ket -
\vert k k' \ket $. 
\cap
According to the definition of Eq. (\re{spe}), Eq. (\re{auxu})
implies an implicit self-consistent procedure.
\par
\begin{figure}[ht]
\vspace{-5 cm}
\begin{center}
\includegraphics[width=1.6\textwidth]{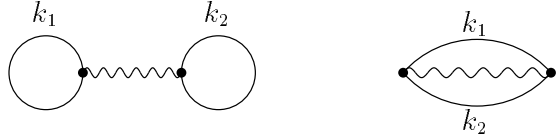}
\end{center}
\vspace{-20.5 cm}
\caption[]{The two hole-line contribution in terms of the Brueckner
G-matrix (wavy line).}
\label{2h}
\end{figure}

Summing up the ladder diagrams to all orders, one then gets the two
diagrams, direct and exchange, of Fig.~\ref{2h}, where a wavy lines
indicates a Brueckner G-matrix. Indeed, if one expands the G-matrix
from Eq. (\re{bruin}), in terms of the bare interaction $v$, and 
inserts the expansion in the diagrams of Fig.~\ref{2h}, one gets the 
full sets of ladder diagrams, indicated in Fig.~\ref{ladder}. 
More details on the rules for 
writing down the explicit expression of the diagrams can be found in
ref. \cite{book}. 
 
\noindent
The first potential insertion diagram, at the bottom of Fig.~\ref{ladder},
cancels out the potential part of the single particle energy of Eq. (\re{spe}),
in the expression for the total energy $E$. This is actually true
for any definition of the auxiliary potential $U$. At the two 
hole-line level of approximation, one therefore gets

\beq
\bar{rl}
   E &= \sum_{k < k_F}  {\hbar^2 k^2\over 2m}
 + {1 \over 2} \sum_{k,k' < k_F}
      \bra k k' \vert G(e_{k}+e_{k'}) \vert k  k' \ket_A \ \ \ \\
&              \\  
 &\equiv \sum_{k < k_F}  {\hbar^2 k^2\over 2m}
  + {1 \over 2} \sum_{k< k_F}  U(k) 
\ear
\le{e2h} 
\eeq
\cap
where, in the last equality, the definition of Eq.~(\re{auxu}) has been
adopted. 
The result that only the unperturbed kinetic energy appears in the
expression for $E$, and all the correlations are included in the
potential energy part, holds true to all orders and it is a peculiarity
of the BBG expansion. Of course, the modification of the momentum 
distribution, and therefore of the kinetic energy, is included in the
interaction energy part, but it is treated on the same footing as the
other correlation effects. This seems to present a noticeable
advantage. In fact, the modification of the kinetic energy in itself
is quite large and, of course, positive and should be therefore 
compensate by an extremely accurate calculations of the 
(negative) correlation energy.
On the other hand, putting the two effects on the same footing, one can 
expect that strong cancellation occur order by order. \par
\begin{figure}[ht]
\vspace{-5 cm}
\begin{center}
\includegraphics[width=1.6\textwidth]{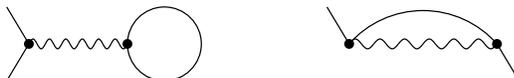}
\end{center}
\vspace{-20.5 cm}
\caption[]{The direct and exchange parts of the auxiliary potential
$U$ in terms of the Brueckner G-matrix.}
\label{upot}
\end{figure}

Let us now discuss the choice of the single particle potential $U$.
As it was discussed in connection with Eq. (\re{split}), the potential
$U$ is in principle arbitrary, and it is used only as a tool for
speed up the convergence of the expansion. However, physical considerations
suggest the self-consistent procedure defined by Eq. (\re{auxu})
to obtain the potential $U$. The self-consistency condition is clearly
non-perturbative and it is a generalization of the usual Hartree-Fock
(HF) approximation, namely the Brueckner G-matrix
is used in place of the bare NN interaction $v$.
 For nuclear matter the HF approximation would produce
unrealistic results, because of the strong repulsive core. The G-matrix
takes into account the short range correlations between pairs of nucleons,
and therefore it gives a much improved balance between attractive and 
repulsive contributions. The approximation of Eq. (\re{e2h}), together
with Eqs. (\re{spe}), (\re{auxu}), is usually referred to as the
Brueckner-Hartree-Fock (BHF) approximation.   
This definition of $U$ corresponds to the
diagrams of Fig.~\ref{upot}.         

\cap
It has to be noticed that the G-matrix appearing in the diagrams are
calculated on-shell, according to Eq. (\re{auxu}), i.e. its entry energy
is equal to the energy of the two particles with the two entry momenta.
Therefore the total energy at the BHF level of approximation can
be written also in terms of the potential $U$, as in the second line
of Eq. (\re{e2h}).\par
In the general BBG expansion, in all the higher order diagrams,
beyond the BHF approximation, the same definition of $U$ is kept and
the bare NN potential is replaced by the G-matrix by performing
the corresponding ladder sums whenever it is possible. In this way
the diagrammatic expansion is rearranged in terms of the Brueckner G-matrix, 
in place of the bare NN interaction,
with the only obvious prescription that no ladder sums can now appear
in the diagrams, just to avoid double counting. \par
\begin{figure}[ht]
\vspace{-3.9 cm}
\begin{center}
\includegraphics[width=1.6\textwidth]{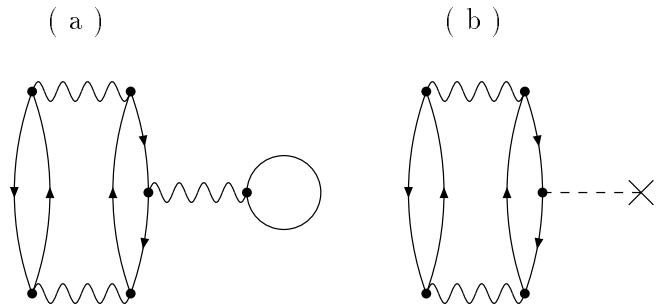}
\end{center}
\vspace{-19.8 cm}
\caption[]{Lowest order three hole-line diagram (a) and the corresponding
potential insertion diagram (b).}
\label{bbp}
\end{figure}

  
We have seen that the ladder sum at the BHF level introduce 
on-shell G-matrices only. 
This is not necessarily the case if the ladder sum is performed
inside a generic higher order energy diagram, since then the entry energy
of the resulting G-matrix depends in general on the rest of the diagrams.
The energy denominators appearing in the BBG expansion include, in fact,
all the particle and hole energies across the diagram. This point
will be discuss later and we will see that some exceptions to this 
expectation can occur.
\cap
Another strong reason in favour of keeping the BHF definition for the single 
particle potential $U$ in the general BBG expansion is the occurrence of
cancellation between diagrams including three hole-lines, thus reducing 
the relevance of higher order contributions. This is true for the two diagrams
reported in Fig.~\ref{bbp}. The diagram ($b$) in the 
right side of the figure is a potential insertion diagram, where the dashed
line with the cross indicates a multiplication by a factor $U(k)$, being
$k$ the momentum of the hole-line to which the potential is attached.
The rule for writing down the potential insertion diagrams can also be found 
in ref. \cite{book}. The diagram ($a$) in the left side of Fig.~\ref{bbp} 
contains a G-matrix loop in place of the potential $U$. If the G-matrix
is on-shell,in view of the definition of
Eq. (\re{auxu}) and the graphical rules, one can easily see that the two 
diagrams cancel out exactly. 
 
\cap
At first site the G-matrix of diagram ($a$) should be not calculated on-shell.
However, it has been shown in ref. \cite{bbp} that, if the ladder sums
included in the diagram contain bare interactions which appear in all possible 
position along the diagram,
then their overall contribution reduces indeed to the diagram ($a$)
of Fig.~\ref{bbp}, with the G-matrix calculated on-shell, and the above
mentioned cancellation holds true.\par
\begin{figure}[ht]
\vspace{-3.4 cm}
\begin{center}
\includegraphics[width=1.6\textwidth]{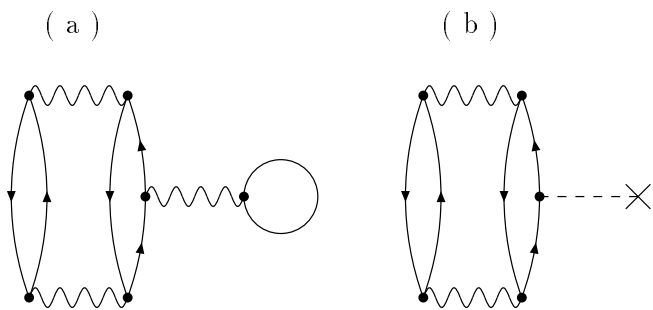}
\end{center}
\vspace{-19.8 cm}
\caption[]{Bubble three hole-line diagram (a) and the corresponding
potential insertion diagram (b)}
\label{bubble}
\end{figure}

The definition of Eq. (\re{auxu}) does not specify completely
the single particle potential $U(k)$. For momenta $k > k_F$ the value
of the potential $U(k)$ does not appear explicitly in the energy expression
of Eq. (\re{e2h}) at the BHF level. In old BHF calculations the potential
$U(k)$ was then taken identically zero above the Fermi momentum, 
with the justification that the interaction between particles above $k_F$
is expected to be small and anyhow only slightly affecting the 
total energy. In this choice, usually referred to as ``standard choice", 
the potential has then a jump at $k_F$. For this reason it is also
often called ``gap choice". Most modern BHF calculations adopt
a potential $U(k)$ which is defined by extending the definition of 
Eq. (\re{auxu}) also above $k_F$, thus making $U(k)$ continuous
across the Fermi sphere. This definition modifies the self-consistent
equation and therefore also the potential for $k < k_F$. As a consequence,
this different choice, usually called ``continuous choice", modifies
indirectly also the value of the BHF energy of Eq. (\re{e2h}).
There are some arguments in favour of the continuous choice. 
Since $U(k)$ has the physical meaning of single particle potential,
it is intimately related to the single particle self-energy. 
Indeed, one can show \cite{mahaux} that $U(k)$ is the on-shell
self energy to first order in the hole expansion. As such, the
potential $U(k)$ must be a continuous function of the momentum.
Another point to be considered is related to the two other 
three hole-line diagrams depicted in Fig.~\ref{bubble}. They 
can be obtained from the diagrams of Fig.~\ref{bbp} just by
attaching the intermediate G-matrix (diagram $a$) and the potential $U$ 
(diagram $b$) to the particle-line instead of the hole-line.
Diagram ($a$) is usually denoted as  ``bubble diagram".
In this case the G-matrix is not calculated on-shell, since the
argumentation of ref. \cite{bbp} does not apply, and no exact cancellation
can occur between the two diagrams. Actually, in the standard choice
the potential insertion diagram $b$ is identically zero, since in this case 
$U(k)$ vanishes for $k > k_F$. On the contrary, in the continuous
choice, the potential insertion diagram does not vanish, and some
degree of cancellation can be expected, despite the G-matrix
is calculated off-shell, thus reducing also in this case the
contribution from higher order diagrams. 

\par
At first sight it can be surprising that the final result for the 
nuclear matter EOS could depend on the choice of the single 
particle potential $U$, since the splitting of Eq. (\re{split})
is a trivial identity and the final result should be independent
on the particular choice of $U(k)$. This is of course true only if the
full BBG expansion to all order could be summed up exactly. 
If the expansion is truncated at a given order, the results 
can show still a dependence on the choice of $U(k)$, and this dependence
will be stronger more the expansion is far from a reasonable 
convergence. One can, therefore, take the degree of the
dependence on $U(k)$ as a measure of the degree of convergence reached
at a given order of the expansion. The gap and continuous choices
can be considered two opposite cases for the potential $U(k)$,
since any other reasonable choice would modify mainly its definition
for $k > k_F$ and would be somehow intermediate between these two cases.
In fact, the exact cancellation between the two diagrams of Fig.~\ref{bbp}
occurs only with the definition of Eq. (\re{auxu}) for $k < k_F$,
and it appears inconvenient to adopt a choice for $U(k)$ which
does not include the cancellation. 
However, other choices are surely possible, and one 
should check also in those other cases the degree of convergence
reached at a given level of the expansion. In the sequel we will
restrict to the gap and continuous choices for checking the convergence
of the expansion.\par
\begin{figure}[ht]
\vspace{-3.5 cm}
\begin{center}
\includegraphics[width=1.\textwidth]{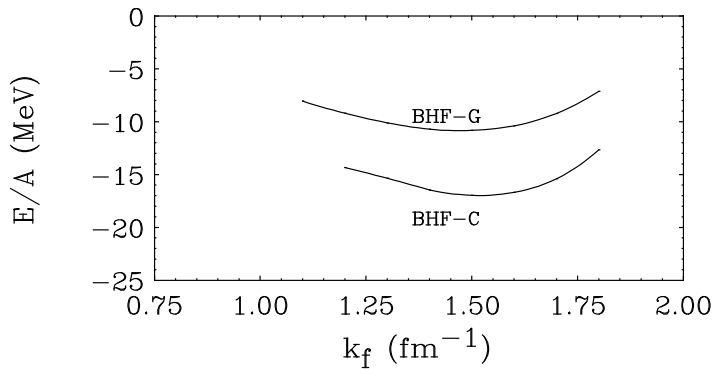}
\end{center}
\vspace{-8.7 cm}
\caption[]{Nuclear matter saturation curve for the Argonne v$_{14}$
 NN potential. The solid lines indicate the results at the 
Brueckner (two hole-lines) level for the standard 
 (BHF-G) and the continuous choices (BHF-C) respectively. }
\label{sat}
\end{figure}

Let us consider the symmetric nuclear matter EOS at the BHF level
of approximation. The results for the two choices for $U(k)$ are reported 
in Fig.~\ref{sat}, where the Argonne v$_{14}$ \cite{v14} is used
for the bare NN potential. 

\cap
It is apparent from the figure that the degree of convergence is not yet
satisfactory at the BHF level. The difference for the energy per particle
is of about 4-5 MeV in the considered density range. It has to be kept 
in mind, however, that the potential energy part of the binding energy of
Eq.~\re{e2h} is about -40 MeV around saturation density, and therefore
the discrepancy between the two choices is of about $10$\%. This is the
expected degree of convergence at the BHF level, according to the above
discussed criterion. 
\par
The BHF results imply that, for a check of convergence, 
it is mandatory to consider the three hole-line diagrams 
contribution. According to the BBG expansion, this set of diagrams
describes the irreducible three-nucleon correlations, i.e. the
three-body correlations which cannot be reduced to a product of two-body
correlations, already introduced at the BHF level.
Let us consider in some detail how the three hole-line diagrams can be 
summed up exactly, in analogy to the summation of the ladder two hole-line
diagrams of the BHF approximation. Indeed, since the two hole-line
contribution has been summed up by introducing the G-matrix, which 
is the in-medium two-body scattering matrix,
it is therefore conceivable that the three hole-line diagrams
could be summed up by introducing some similar generalization
of the scattering matrix for three particles.
The three-body scattering problem
has a long history by itself, and has been given a formal
solution by Fadeev \cite{Fadeev}. For three distinguishable particles
the three-body scattering matrix $T^{(3)}$ is expressed as the
sum of three other scattering matrices, $T^{(3)} = T_1 +
T_2 + T_3$. The scattering matrices $T_i$ satisfy a system
of three coupled integral equations. The kernel of this 
set of integral equations contains explicitly the two-body
scattering matrices pertaining to each possible pair of 
particles. Also in this case, therefore, the original 
two-particle interaction disappears from the equations in 
favour of the two-body scattering matrix. For identical particles the 
three integral equations reduce to one, because of symmetry.
In fact, the three functions $T_i$ must coincide within a
change of variable with a unique function, which we can still
call $T^{(3)}$. The analogous equation and scattering matrix in 
the case of nuclear matter (or other many-body systems in
general) has been introduced by Bethe \cite{Bethe,Raja}. The integral
equation, the Bethe--Fadeev equation, reads schematically
\beq
\bar{l}
 T^{(3)} \,\, =\,\,  G\, \,+\,\, G\,\, X\,\, {Q_3 \over e}\,\, T^{(3)} \\
                        \\
 \bra k_1 k_2 k_3 \vert T^{(3)} \vert k'_1 k'_2 k'_3 \ket
  \,=\,  \bra k_1 k_2 \vert G \vert k'_1 k'_2 \ket 
  \delta_K (k_3 - k'_3) \,+\, \\ 
            \\
\phantom{ 
\bra k_1 k_2 k_3 \vert T^{(3)} \vert k'_1 k'_2 k'_3 \ket \,=\, } 
\,+\,  \bra k_1 k_2 k_3 \vert G_{12}\, X\, {Q_3 \over e}\, T^{(3)} 
   \vert k'_1 k'_2 k'_3 \ket   \ \ \ .  \\
\ear
\le{fads}
\eeq
\cap                                                  
The factor $Q_3 /e$ is the analogous of the similar factor
appearing in the integral equation for the two-body 
scattering matrix $G$, see Eq. (\re{bruin}). Therefore, the projection
operator $Q_3$ imposes that all the three particle states lie above the 
Fermi energy, and the denominator $e$ is the appropriate energy
denominator, namely the energy of the three-particle intermediate
state minus the entry energy $\omega$, in close analogy with
the equation for the two-body scattering matrix $G$ of Eq. (\re{bruin}).
The real novelty
with respect to the two-body case is the operator $X$.
This operator interchanges particle $3$ with
particle $1$ and with particle $2$, $X = P_{123} + P_{132}$,
where $P$ indicates the operation of cyclic permutation of 
its indices. It gives rise to the so-called ``endemic factor'' 
in the Fadeev equations, since it is an unavoidable complication 
intrinsic to the three-body problem in general. The reason for the
appearance of the operator $X$ in this context is that no two successive 
$G$ matrices can be present in the same pair of particle
lines, since the $G$ matrix already sums up all the two-body ladder
processes. In other words, the $G$ matrices must alternate from one pair
of particle lines to another, in all possible ways, as it is indeed
apparent from the expansion by iteration of Eq. (\re{fads}), 
which is represented in Fig.~\re{ladfad}.
\begin{figure}[ht]
\vspace{-6. cm}
\begin{center}
\includegraphics[width=1.6\textwidth]{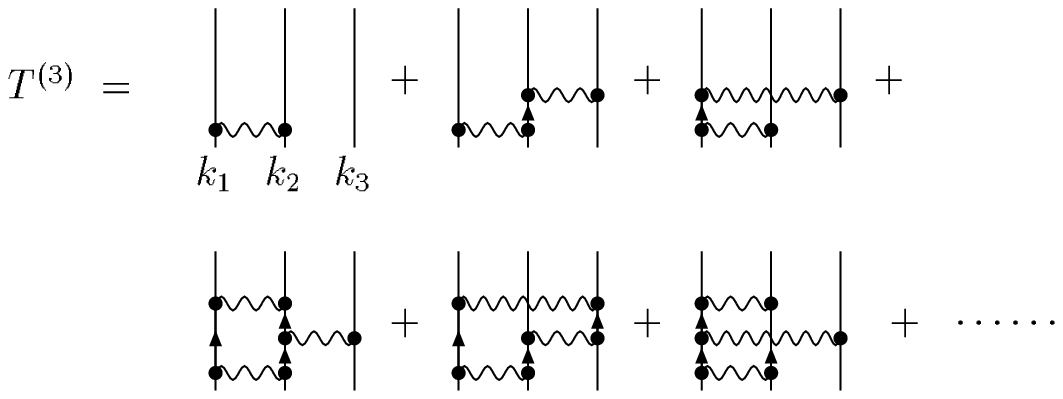}
\end{center}
\vspace{-17.2 cm}
\caption[]{The first few terms in the expansion of the 
 Bethe-Fadeev integral equation.}
\le{ladfad}
\end{figure}
\par\noindent
Therefore, both cyclic operations are necessary in order to include
all possible processes. Adding all terms with an arbitrary number
of G-matrices, one gets a generalized ladder series for three-particles,
analogous to the ladder series introduced for the two particles case 
in defining the G-matrix. Indeed, this is the basis for the
integral equation (\re{fads}).
In the structure of Eq. (\re{fads}) the third particle, 
with initial momentum $k_3$, is somehow singled out from the other two.
This choice is arbitrary, but it is done in view of the use of
the Bethe-Fadeev equation within the BBG expansion.
\par
\begin{figure}[ht]
\vspace{-3.4 cm}
\begin{center}
\includegraphics[width=1.6\textwidth]{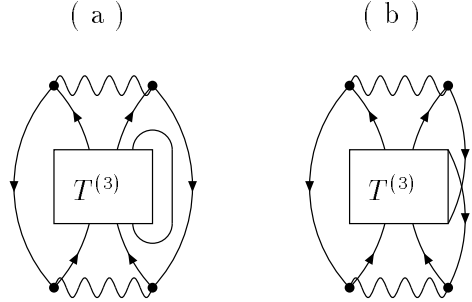}
\end{center}
\vspace{-18.8 cm}
\caption[]{Schematic representation of the direct (a) and
 exchange (b) three hole-line diagrams.}
\label{3h}
\end{figure}

In order to see how the introduction of the three-body scattering
matrix $T^{(3)}$ allows to sum up the three hole-line diagrams,
we first notice, following B.D.Day \cite{Day1980}, that this set of diagrams
can be divided into two distinct groups. The first one includes
the graphs where two hole-lines, out of three, originate at the
first interaction of the graph and terminate at the last one 
without any further interaction in between. Schematically 
the sum of this group of diagram can be represented as in part (a) 
of Fig.~\ref{3h}.
\cap
The third hole-line has been explicitly indicated, out from
the rest of the diagram. The remaining
part of the diagram describes the rescattering, in all possible
way, of three particle-lines, since no further hole-line must
be present in the diagram. This part of the diagram is indeed
the three-body scattering matrix $T^{(3)}$, and the operator
$Q_3$ in Eq. (\re{fads}) assures, as already mentioned, that only particle 
lines are included.
\par
\begin{figure}[ht]
\vspace{-5. cm}
\begin{center}
\includegraphics[width=1.6\textwidth]{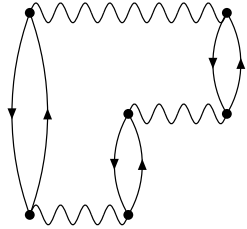}
\end{center}
\vspace{-19.3 cm}
\caption[]{The ring diagram, belonging to the set of three hole-line 
 diagrams. It can be considered the exchange diagram of the bubble
 diagram.}
\label{ring}
\end{figure}

The second group includes the diagrams where two of the 
hole-lines enter their second interaction at two different vertices 
in the diagram, as represented in part (b) of Fig~\ref{3h}.
Again the remaining part of the diagram is $T^{(3)}$, i.e. the sum of
the amplitudes for all possible rescattering process of three particles. 
It is easily seen that no other structure is possible. 
The set of diagrams indicated
in part (b) can be obtained by the ones of part (a) by simply
interchanging the final (or initial) point of one of the
``undisturbed'' hole-line with the final (or initial) point of
the third hole-line. This means that one can obtain each graph
of the group depicted in Fig.~\ref{3h}b  by acting with the operator 
$X$ on the bottom of the corresponding graph of Fig.~\ref{3h}a. In this
sense the diagrams of Fig.~\ref{3h}b can be considered the ``exchange''
diagrams of the ones in Fig.~\ref{3h}a (not to be confused with the
term ``exchange'' introduced previously for the matrix elements
of $G$). If one inserts the terms obtained by iterating Eq. (\re{fads})
inside these diagrams in substitution of the scattering matrix $T^{(3)}$
(the box in Fig.~\ref{3h}), the first diagram, coming from the inhomogeneous 
term in Eq. (\re{fads}), is just the bubble diagram of Fig.~\ref{bubble}.
The corresponding exchange diagrams is the so called ``ring diagram"
of Fig.~\ref{ring}.
\cap
It is easy to draw the remaining series of diagram which one obtains
by going on with the iterations. 
\par
Once the Bethe-Fadeev equations are solved, the contribution of the
direct three hole-line diagrams of Fig.~\ref{3h}a can be written as
\beq
\bar{c}
 E^{dir}_{3h} = {1 \over 2} \sum_{k_1,k_2,k_3 \leq k_F}
\sum_{\{k'\}, \{k''\} \geq k_F}
 \bra k_1 k_2 \vert G \vert k'_1 k'_2 \ket_A  \cdot \\
                 \\
\ \ \ \ \ \cdot {1\over e}\,\, \bra k'_1 k'_2 k'_3 \vert X T^{(3)} X 
\vert k''_1 k''_2 k''_3 \ket\,\, {1 \over e'}\,\,
   \bra k''_1 k''_2 \vert G \vert k_1 k_2 \ket_A  \ \ \ ,
\ear
\le{efad}
\eeq
\cap
In Eq. (\re{efad})
the denominator $e = e_{k'_1} + e_{k'_2} - e_{k_1} - e_{k_2}$, 
and analogously $e' = e_{k''_1} + e_{k''_2} - e_{k_1} - e_{k_2}$.
The exchange diagrams of Fig. \ref{3h}b can be obtained by multiplying
the same expression by a further factor $X$.
 In summary, the entire set of three hole-line diagrams 
can be obtained by multiplying the expression of Eq. (\re{efad}) by $1 + X$.
\par
It has been recognized a long ago \cite{Raja} that the summation
of all three-hole diagrams is essential, since individual three-hole
diagram can be quite large, but strong cancellation occurs among
the different contributions. This is particularly true for the
bubble diagram of Fig. \ref{bubble}a and the ring diagram of
Fig. \ref{ring}, which turn out to be quite large but of opposite
sign. As already mentioned, the potential insertion diagram
of Fig. \ref{bubble}b is different from zero in the continuous
choice and it turns out to be essential in compensating the
contribution of both bubble and ring diagrams. 
A scheme of approximation was first devised by B.D. Day \cite{Day1980}
within the gap choice for the single particle potential.
In this scheme the bubble and ring diagrams are indeed singled out from the
whole set of three hole-line diagrams, while the remaining series of 
diagrams is summed up by solving the Bethe-Fadeev integral equation.
The bubble diagram requires special numerical treatment, since very
large partial waves contribute to the intermediate G-matrix. 
Once the bubble and ring diagrams are subtracted from the
Bethe-Fadeev equation, the resulting integral equation for the whole
set of the higher order diagrams 
 turns out to be much less sensitive to the 
larger partial waves. We will refer to this contribution as the
``higher order" contribution. The numerical solution of the
Bethe-Fadeev integral equation is delicate. The main difficulty
is the large matrix to be inverted to get the scattering matrix
$T^{(3)}$. This difficulty can be overcome by introducing a
separable representation of the G-matrix appearing in the kernel
of the integral equation, as already performed by B.D. Day \cite{Day1980}
in the case of the gap choice. We refer to this reference and to
ref. \cite{book} for other details of the numerical methods. \par
\begin{figure}[ht]
\vspace{-1.5 cm}
\begin{center}
\includegraphics[width=0.95\textwidth]{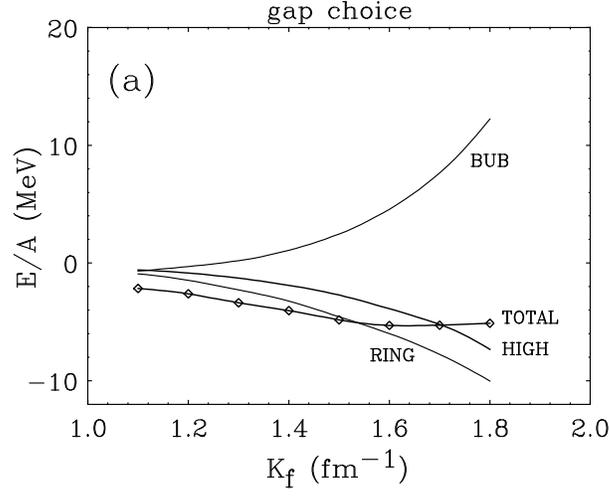}
\end{center}
\vspace{-8.4 cm}
\caption[]{The contributions of the bubble (BUB) , ring (RING) and 
higher order (HIGH) diagrams 
to the binding energy of symmetric nuclear matter as a function of
Fermi momentum, calculated within the gap choice. 
The line denoted by TOTAL is the sum of all these contributions
and gives the overall three hole-line contribution to the EOS.}
\label{fad3a}
\end{figure}
\begin{figure}[ht]
\vspace{-1.5 cm}
\begin{center}
\includegraphics[width=0.95\textwidth]{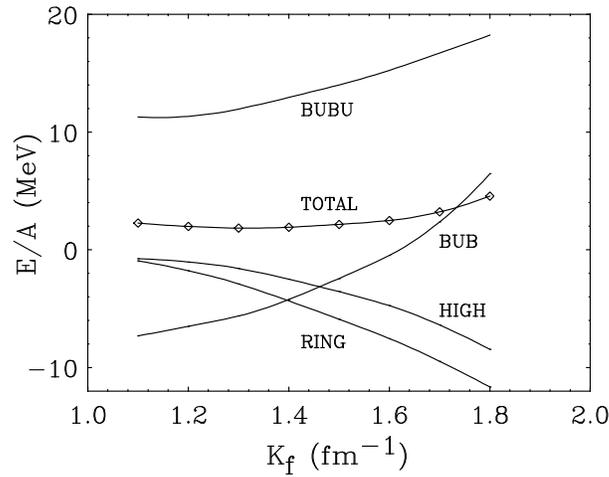}
\end{center}
\vspace{-8.4 cm}
\caption[]{The same as in Fig.~\ref{fad3a}, but within the continuous choice.
Here the line denoted by BUBU is the contribution of the potential
insertion diagram of Fig.~\ref{bubble}b.}
\label{fad3b}
\end{figure}

The degree of cancellation among the different terms is apparent
in Fig.~\ref{fad3a}, where the bubble, ring and higher order
contributions are displayed \cite{song} in the case of the gap choice 
and the Argonne v$_{14}$ NN potential. 
\cap
The final result, denoted
as ``total", is relatively small and much smaller in size than the
individual contributions. The corresponding results 
for the continuous choice are displayed in Fig.~\ref{fad3b}. In this
case the additional contribution (BUBU) of the potential insertion
diagram in Fig.~\ref{bubble}b must be considered. One can see the relevance 
of this term in comparison with the others and its role in determining
the size of the total three hole-line contribution. The latter turns out to be
much smaller in the continuous choice than in the gap choice. \par
\begin{figure}[ht]
\vspace{-3.5 cm}
\begin{center}
\includegraphics[width=1.\textwidth]{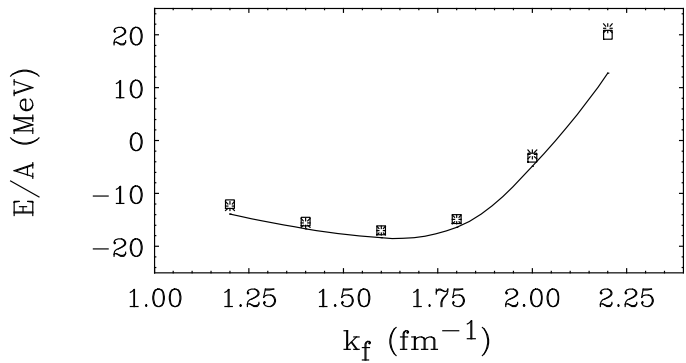}
\end{center}
\vspace{-8.7 cm}
\caption[]{The Nuclear Equation of State including the three hole-line
contribution within the
gap choice (squares) and the continuous choice (stars),  for the
Argonne v$_{14}$ potential. For comparison, the EOS at the two hole-line
level in the continuous choice is also reported (solid line).}
\label{fad_song}
\end{figure}

\cap
The final Equation of State obtained by adding the three hole-line
contribution is reported in Fig.~\ref{fad_song}, both for the
gap choice (squares) and the continuous choice (stars), again for the
Argonne v$_{14}$ potential, for a much wider range of densities than
in Fig.~\ref{sat}. For comparison, the EOS at the two hole-line
level in the continuous choice is also again reported (solid line)
from Fig.~\ref{sat}. Two conclusions can be drawn from these results.
\cap
i)\  The two saturation curves in the gap and continuous choices, 
with the inclusion of the three hole-line diagrams, tend now to collapse
in a single EOS, with some deviations only at the highest density.
This is a strong indication that a high degree of convergence has been
reached at this level of the expansion, according to the criterion
discussed above.
Notice that the saturation curves extend from a density which is about 
one half of saturation density to about five times saturation density,
and, therefore, it appears unlikely that the agreement between
the two choices can be considered as a fortuitous coincidence.
\cap
ii)  The Brueckner two hole-line EOS within the continuous choice
turns out to be already close to the full EOS, since in this case
the three hole-line contribution is quite small. In first approximation
one can adopt the BHF results with the continuous choice as the nuclear
matter EOS.
\cap
The phenomenological saturation point for symmetric nuclear matter is, 
however, not reproduced, which confirms the finding in ref. \cite{Day1980}. 
The binding energy per particle at the minimum
of the saturation curve turns out to be
close to the empirical value of about -16 MeV, but the corresponding
density comes out about 30-40 $\%$ larger than the empirical one.
Usually this drawback is corrected by introducing three-body forces
in the nuclear hamiltonian, and indeed all realistic two-nucleon 
forces, which fit the experimental two-nucleon phase shifts 
and deuteron data, are not able to reproduce the empirical saturation point. 
In other words, the results indicate that the missing of the saturation 
point is not due to a lack of accuracy in the treatment of the nuclear 
many-body problem, but to a defect of the nuclear hamiltonian.  
The need of three-body forces in nuclear matter is consistent
with the findings in the study of few nucleon systems, where also 
the binding energy and radii, as well as scattering data,
cannot be reproduced with only two-body forces. Not surprisingly,
the effects of three-body forces seem to be more pronounced in
nuclear matter than in few body systems.
\par
\begin{figure}[ht]
\vspace{-4.0 cm}
\begin{center}
\includegraphics[width=1.6\textwidth]{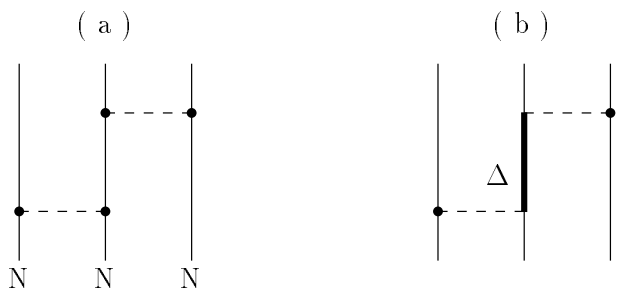}
\end{center}
\vspace{-19.4 cm}
\caption[]{An interaction process among three nucleons with only
two-body force (a), and
 a process involving a genuine three-body force (b).}
\label{delta}
\end{figure}
\begin{figure}[ht]
\vspace{-4.5 cm}
\begin{center}
\includegraphics[width=1.6\textwidth]{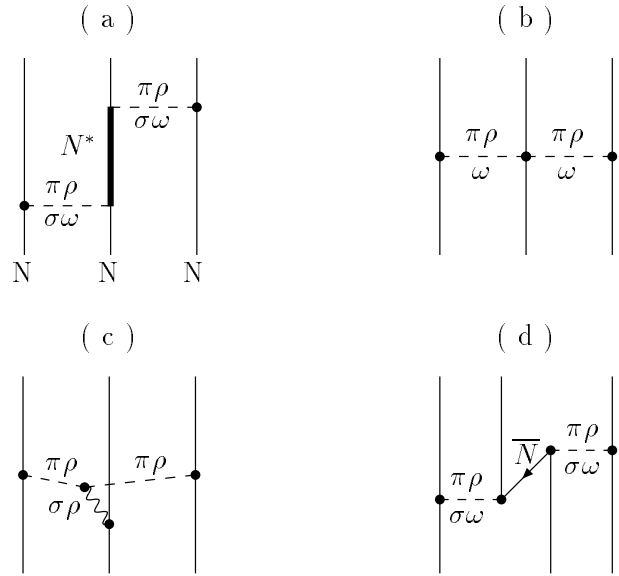}
\end{center}
\vspace{-15.2 cm}
\caption[]{Some of the processes which can produce a genuine
three-body force.}
\label{tbf}
\end{figure}

The standard NN interaction models are based
on the meson--nucleon field theory, where the nucleon is
considered an unstructured point-like particle.
The Paris, the Argonne $v_{14}$ (with the improved
version $v_{18}$ \cite{v18}), and the set of Bonn potentials \cite{Bonn} 
fall in this category. In the one-boson exchange potential (OBEP)
model one further assumes that no meson--meson interaction is
present and each meson is exchanged in a different interval of time
from the others. However, the nucleon is a structured particle, it
is a bound state of three quarks with a gluon-mediated interaction,
according to Quantum Chromodynamics (QCD). The absorption and emission 
of mesons can be accompanied by a modification of the nucleon
structure in the intermediate states, even in the case of NN scattering
processes, in which only nucleonic degrees of freedom are present
asymptotically. A way of describing such processes is to introduce
the possibility that the nucleon can be excited (``polarized'') to other
states or resonances. The latter can be the known resonances 
observed in meson--nucleon scattering. At low enough energy the
dominant resonance is the $\Delta_{33}$, which is the lowest in mass.
If the internal nucleon state can be distorted by the presence of another
nucleon, the interaction between two nucleons is surely altered by the
presence of a third one. This effect produces clearly a definite
three-body force, which is absent if the nucleons are considered
unstructured. The simplest of such process is depicted in Fig.~\ref{delta}b.
\cap
Such a process can be interpreted in different but equivalent ways.
One way is to view the pion (meson) coming from the first nucleon to 
polarize the second one, which therefore interacts with a third one as a
$\Delta_{33}$ resonance, surely in a different way than if it had remained 
a nucleon, like in Fig.~\ref{delta}a. The process of Fig.~\ref{delta}a
is not indeed a 
three-nucleon force, but just a repetition of a two-nucleon force.
The introduction of a three-nucleon interaction is a consequence of
viewing processes like the one of Fig.~\ref{delta}b as an effective interaction 
among three nucleons, which eventually will be medium-dependent.
The genuine three-nucleon forces
can be extracted from processes like the one of Fig.~\ref{delta}b by
projecting out the $\Delta_{33}$ (or other resonances) degrees of freedom 
in some approximate way. 
The theory of three-nucleon forces has
a very long history, and it started to be developed since the early stage
\cite{earl3bf} of the theory of nuclear matter EOS, as well as of few nucleon 
systems \cite{Sauer}. The most extensive study of the three-nucleon forces
(TNF) has been pursued by Grang\'e and collaborators \cite{Gran}. 
Fig.~\ref{tbf}, 
reproduced from Ref. \cite{Math}, indicates some of the processes which can 
give rise  to TNF. Graph of Fig.~\ref{tbf}a is a generalization of the 
process of
Fig.~\ref{delta}b, where other nucleon resonances (e.g. the Roper resonance)
can appear as intermediate virtual excitation and other 
exchanged mesons can be present.
Graph ~\ref{tbf}b includes possible non-linear meson-nucleon coupling,
as demanded by the chiral symmetry limit \cite{Math}.
Graph ~\ref{tbf}c is the simplest one which includes meson-meson 
interaction. Other processes of this type are of course 
possible \cite{Gran,Math}, which involves other meson-meson couplings,
and they should be included in a complete treatment of TNF. 
Diagram ~\ref{tbf}d describes the effect of the virtual
excitation of a nucleon-antinucleon pair, and it is therefore somehow
of different nature from the others. It gives an important (repulsive)
contribution and it has been shown \cite{BrownWeise} to describe the 
relativistic effect on the EOS to first order in the ratio $U/m$ between the
single particle potential and the nucleon rest mass.\par
The $\sigma$ meson, appearing in some of the diagrams, is a hypothetical
scalar meson, believed to be responsible for the intermediate
attraction in the two-nucleon interaction, whose mass and coupling
constant are treated as parameters. One should 
therefore be careful, as discussed in Ref. \cite{Gran}, to be at least 
consistent between the treatments of the two-nucleon and the three-nucleon 
forces. A complete calculations of the TNF in the framework of the 
meson-nucleon theory, i.e. the calculation of the ``best" TNF, is not yet 
available.\par
A simpler possibility is to
adopt a more phenomenological approach, like the one
followed by the Urbana group \cite{3bfUrb}. 
Since the EOS obtained with only two-body forces seems
to need additional attraction at lower density and an additional
repulsion at higher density, it is therefore conceivable that the
main effect of TNF can be schematized by one attractive and one
repulsive term, as representative of the whole set of three-nucleon
processes. 
Actually, once the usual static approximation is made for the
nucleons and the resonances in calculating the meson exchange
process, the structure of the different
three-body forces turns out to be quite similar. Since the 
strengths of the different vertex appearing in these diagrams
cannot be considered fairly well known, one can treat the strengths
of the two representative terms as free parameters to be fitted
to some known physical quantities. More explicitly, the TNF is written as
\beq
\bar{c} 
 V_{ijk} =  V^{2\pi}_{ijk} +  V^{R}_{ijk} \ \ \ .
\ear
\le{tnf1}
\eeq
\noindent
The first (attractive) contribution is a cyclic sum over the nucleon indices 
{\it i, j, k} of products of anticommutator \{,\} and commutator [,] terms
\beq
\bar{rl}
V^{2\pi}_{ijk} & =   A \sum_{cyc} \Big( \{X_{ij},X_{jk}\} 
 \{\tau_i \cdot \tau_j,\tau_j \cdot \tau_k\}  \\
 &                \\
& +  {{1}\over {4}} 
  [X_{ij},X_{jk}] [\tau_i \cdot \tau_j,\tau_j \cdot \tau_k] \Big) \ \ ,
\ear
\le{tnf2}
\eeq
\noindent
where
\beq
\bar{c}
X_{ij} = Y(r_{ij}) \sigma_i \cdot \sigma_j + T(r_{ij}) S_{ij} 
\ear
\eeq
\noindent
is the one--pion exchange operator, $\sigma$ and $\tau$ are the Pauli spin 
and isospin operators, and 
$ S_{ij} =  3 \big[ (\sigma_i \cdot r_{ij})(\sigma_j \cdot r_{ij})
                - \sigma_i \sigma_j \big ] $   
is the  tensor operator. 
$Y(r)$ and $T(r)$ are the Yukawa and tensor functions, respectively, 
associated to the one--pion exchange, as in the two--body potential.
 The repulsive part is taken as  
\beq
\bar{c}
V^{R}_{ijk} = U \sum_{cyc} T^2(r_{ij})  T^2(r_{jk}) \ \ \ .
\ear
\le{tnf3}
\eeq
\noindent
\cap
The strengths $A$ ($< 0$) and $U$ ($> 0$) can be fitted to reproduce
the ground state energy of both three nucleon systems (triton and 
$^3He$), and the empirical nuclear matter saturation point. 
\begin{figure} [ht]
 \begin{center}
\includegraphics[bb= 90 130 515 694,angle=90,scale=0.5]{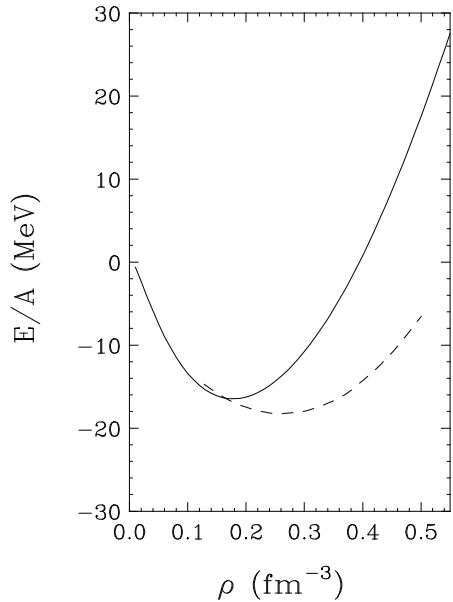}
\end{center}
\caption[]{Saturation curve for symmetric nuclear matter
 in the Brueckner approximation without (dashed curve) and with
 (full line) three-body forces.}
    \label{eostbf}
\end{figure}
\cap
 One such a fit, within the
Brueckner approximation, is reported in Fig.~\ref{eostbf}. The empirical
saturation point is now reproduced and the EOS become much more repulsive
at high density.
Of course,
the higher density region, needed e.g. in  neutron star studies, 
is obtained by extrapolating the TBF from the region around
 saturation 
where they are actually adjusted. This EOS can
therefore be inaccurate at the higher densities. One can see indeed
that the contribution of the three-body forces is substantial
at high density, and therefore an accurate inclusion of the three-body
forces is highly demanded.\par
More detail on the use of phenomenological three-body forces will be
given in the Section on neutron stars.

\section{The EOS for pure neutron matter}

In this Section we will
extend the analysis to pure neutron matter EOS, which is more appropriate
for neutron star studies, at densities up to about five times the 
saturation one. Moreover, we consider the calculations for two 
nucleon-nucleon  potentials, the Av$_{14}$  and the Av$_{18}$,
in order to analyse the dependence of the results on the nuclear
interaction.
\par
\begin{figure} [ht]
\vspace{0.5 cm}
 \begin{center}
\includegraphics[bb= 90 130 515 694,angle=90,scale=0.5]{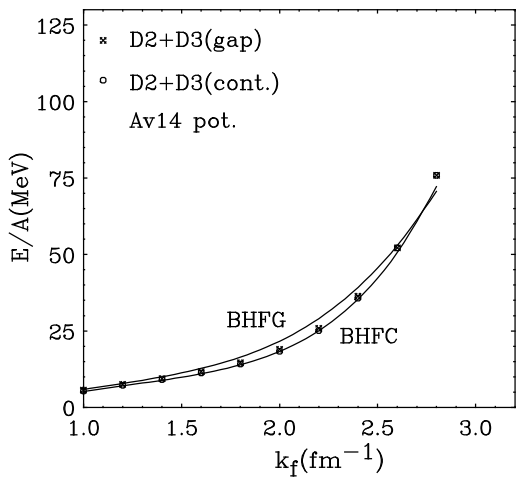}
\end{center}
\vspace{-1.2 cm}
\caption[]{Equation of state of pure neutron matter for the Av$_{14}$
nucleon-nucleon potential. The two full lines correspond to the 
Brueckner-Hartree-Fock approximation, in the gap (BHFG) and continuous
choice (BHFC) respectively. The addition of the three-hole contribution
D$_3$ gives the total equation of state for the gap (stars) and 
continuous choice (open circles) respectively.}
    \label{neumat1}
\end{figure}
\begin{figure} [ht]
\vspace{0.5 cm}
 \begin{center}
\includegraphics[bb= 90 130 515 694,angle=90,scale=0.5]{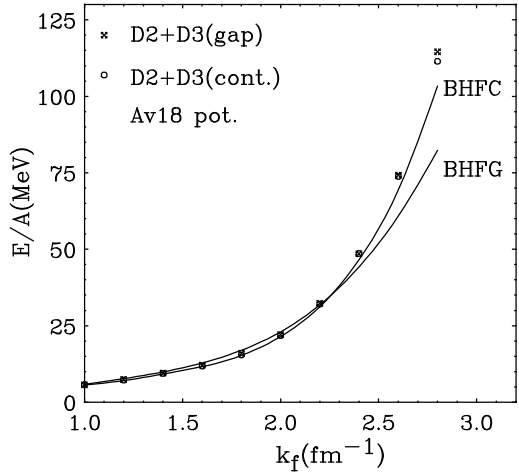}
\end{center}
\vspace{-1.2 cm}
\caption[]{The same as in Fig.~\ref{neumat1}, but for the Argonne v$_{18}$
 potential}
    \label{neumat2}
\end{figure}

We will not give the detail about the contributions of different
diagrams, but simply illustrate the results for the neutron EOS,
obtained by including only two-body forces. 
The neutron matter EOS \cite{neumat} is reported (full lines) 
in Figs.~\ref{neumat1} and \ref{neumat2},
both for the continuous choice (BHFC) and standard choice (BHFG).
As for symmetric nuclear matter, the discrepancy between the two 
curves indicates
to what extent the EOS still depends on the choice of the
auxiliary potential at BHF level, and therefore the degree of convergence.
The EOS for the Av$_{18}$ appears
more repulsive, but the trend for the two potentials is similar.
The discrepancy does not exceed 4 MeV in the whole density
range for the Av$_{14}$ potential, 
and it is tiny in the case of Av$_{18}$, except for the highest
densities. It is also
substantially smaller than in the symmetric nuclear matter case,
where the discrepancy is large as much as about 8 MeV 
at $k_F = 1.8 fm^{-1}$ for the Av$_{14}$ . 
According to our criterion, this suggests
a smaller value of the three hole-line contribution and is in agreement with
the smaller value in neutron matter of the ``wound parameter", which 
is the smallness parameter of the expansion and should give a rough
estimate of the ratio between the three hole-line and the two hole-line
contributions (in general between two successive order contributions).
It can be estimated by the average depletion of the momentum distribution
below the Fermi momentum.
Indeed, at densities around the saturation value the wound parameter
turns out to be close to 0.1 in neutron matter \cite{umb1} and about 0.25
in symmetric nuclear matter \cite{umb2}.
\par
These expectations are indeed confirmed
by the calculations of the three hole-line contributions. 
The inclusion of the three hole-line contributions results in the two
final EOS depicted in Figs.~\ref{neumat1} and \ref{neumat2}, 
where the points marked by stars 
and the circles correspond to the standard and continuous choices,
respectively. For both NN potentials, the very close agreement between the 
two EOS is a strong evidence that the expansion has reached convergence. 
Notice that, at Brueckner (two hole) level, the EOS for the standard and 
continuous choices cross at some value of the density, and at that point the 
overall three hole-line contribution has the same value in both choices.
Furthermore, in the continuous choice the three hole-line contribution is 
substantially smaller, and it is actually negligible to a first approximation. 
It appears that the corresponding values of the wound parameter, which
is close to 0.1 in neutron matter \cite{umb1}, give an upper 
limit for the ratio between three and two hole-line contributions. 
\par
The final EOS appears more repulsive at high density for the Av$_{18}$ than 
for the Av$_{14}$ potential. At lower density, up to about $k_F = 2.0 fm^{-1}$,
the two potentials produce very close EOS. This is not surprising, since
both potentials fit the NN experimental phase shifts up to 350 MeV Lab
energy, which indeed corresponds to a relative momentum of about 
$ 2.0 fm^{-1}$. Above this density, the main contribution to the EOS 
comes from values of the relative NN momentum which need extrapolation beyond 
the region where the potentials have been fitted to the empirical data. 
It is likely that different extrapolations are obtained from 
different potentials in general, and therefore the EOS at high density
is largely dependent on the NN potential model, even without the inclusion
of three-body forces. The inclusion of three-body forces in pure neutron 
matter is discussed in the next Section.

\section{Neutron Star Structure}

Once the Nuclear Equation of State, both for symmetric and pure neutron
matter, has been established on a firm basis, one can try to study
the structure of neutron star (NS) interior. It is indeed believed that the
interior of neutron stars is mainly formed by asymmetric nuclear matter
with increasing density towards the center. The exterior crust should be 
actually formed by different state of matter at sub-nuclear density, 
namely crystal structures of atomic species of increasing mass number.
In the region where the nuclei start to drip, the crystalline structure is
probably mixed with a neutron gas, until nuclei merge into uniform
asymmetric nuclear matter. This outer region is the place where many 
interesting phenomena occur. However, to the extent that  
the analysis is restricted to the mass and radius of the star, 
the main contribution is
coming from the interior, where nuclear matter sets in. One can hope,
therefore, that neutron stars could be a testing ground for the Nuclear
Equation of State. The neutron star masses are usually affected by
large uncertainties and independent measurements of the radii 
have not yet been performed. Only recently some indirect 
indications of neutron star radii have been reported, 
and, as already noticed, the
astrophysics of neutron stars is rapidly developing. An accurate
enough measurement of both mass and radius of a neutron star
is expected to produce an enormous advancement in our knowledge
in the nuclear Equation of State.\par
The observed neutron star masses are $\approx (1-2) M_{\odot}$
(where $M_{\odot}$ is the mass of the sun, $M_\odot = 1.99\times 10^{33}$g).
Typical radii of NS are thought to be of order 10 km, 
and the central density is a few times 
normal nuclear matter density ($\rho_0\approx 0.17\;{\rm fm}^{-3}$). 
This requires a detailed knowledge of the EOS for densities $\rho \gg \rho_0$.
This is a very hard task from the theoretical point of view. 
In fact, whereas at densities $\rho \approx \rho_0$ the matter consists 
mainly of nucleons and leptons, at higher densities several species 
of particles may appear due to the fast rise of the baryon chemical potentials 
with density. Among these new particles are strange baryons, namely, 
the $\Lambda$, $\Sigma$ and $\Xi$ hyperons. 
Due to its negative charge, the $\Sigma^-$ hyperon is the 
first strange baryon expected to appear with increasing density in the 
reaction $n+n \rightarrow p+\Sigma^-$,
in spite of its substantially larger mass compared to the neutral 
$\Lambda$ hyperon ($M_{\Sigma^-}=1197\;{\rm MeV}, M_\Lambda=1116\;{\rm MeV}$).
Other species in stellar matter may appear,
like $\Delta$ isobars along with pion and kaon condensations.
It is therefore mandatory to generalize the study of nuclear EOS
with the inclusion of the possible hadrons, other than nucleons, which
can spontaneously appear in the inner part of a NS, just because their
appearance is able to lower the ground state energy of the nuclear matter
dense phase. In the following we will concentrate  on the production 
of strange baryons and assume that a baryonic description of nuclear matter 
holds up to densities as those encountered in the core of neutron stars. 
\par
As we have seen from the previous Sections, the nuclear EOS can be 
calculated with good accuracy in the Brueckner two hole-line 
approximation within the continuous choice for the single particle
potential, since the results in this scheme are quite close to the full
convergent calculations which include also the three hole-line
contribution. It is then natural to include the hyperon degrees of freedom
within the same approximation to calculate the nuclear EOS needed
to describe the NS interior. To this purpose, one needs also
a nucleon-hyperon (NY) and a hyperon-hyperon (YY)interaction. In the
following this interaction will be 
taken as the Nijmegen soft-core model \cite{mae89}. 
In the calculations the hyperon-hyperon interaction will be neglected
in first approximation. We will comment on this point in the sequel. 
With these NN and NY potentials, the various $G$ matrices are evaluated 
by solving numerically the Brueckner equation, which can be written in 
operatorial form as
\begin{equation}
  G_{ab}[W] = V_{ab} + \sum_c \sum_{p,p'} 
  V_{ac} \Big|pp'\Big\rangle 
  { Q_c \over W - E_c +i\epsilon} 
  \Big\langle pp'\Big| G_{cb}[W] \:, 
\label{e:g}
\end{equation}
where the indices $a,b,c$ indicate pairs of baryons
and the  Pauli operator $Q$ and energy $E$ 
determine the propagation of intermediate baryon pairs.
In a given nucleon-hyperon channels $c=(NY)$ one has, for example,
\begin{eqnarray}
  E_{(NY)} &=& m_N + m_Y + {k_N^2\over2m_N} + {k_Y^2\over 2m_Y} +
  U_N(k_N) + U_Y(k_Y) \:.
\label{e:e}
\end{eqnarray}
The hyperon single-particle potentials within the continuous choice
are given by
\begin{eqnarray}
  U_Y(k) &=& {\rm Re}\, \sum_{N=n,p}\sum_{k'<k_F^{(N)}} 
  \Big\langle k k' \Big| G_{(NY)(NY)}\left[E_{(NY)}(k,k')\right] 
  \Big| k k' \Big\rangle 
\label{e:uy}
\end{eqnarray}
and similar expressions of the form
\begin{eqnarray}
  U_N(k) &=& \sum_{N'=n,p} U_N^{(N')}(k) + 
\sum_{Y=\Sigma^-,\Lambda} U_N^{(Y)}(k) 
\label{e:un}
\end{eqnarray}
apply to the nucleon single-particle potentials.
The nucleons feel therefore direct effects of the other nucleons as well as 
of the hyperons in the environment, whereas for the hyperons there are only 
nucleonic contributions, because of the missing hyperon-hyperon potentials.
The equations (\ref{e:g}--\ref{e:un}) define the BHF scheme with the 
continuous choice of the single-particle energies.  
Due to the occurrence of $U_N$ and $U_Y$ in Eq.~(\ref{e:e}) they constitute 
a coupled system that has to be solved in a self-consistent manner
for several Fermi momenta of the particles involved. 
Once the different single-particle potentials are known,
the total nonrelativistic baryonic energy density, $\epsilon$,  
and the total binding energy per baryon, $B/A$, can be evaluated 
\begin{eqnarray}
 {B\over A} &=& {\epsilon\over \rho_n+\rho_p+\rho_{\Sigma^-}
+\rho_\Lambda} \:,
\\
 \epsilon &=& \sum_{i=n,p,\Sigma^-,\Lambda} \int_0^{k_F^{(i)}}\!\! 
{dk\,k^2\over\pi^2} 
 \left( m_i + {k^2\over{2m_i}} + {1\over2}U_i(k) \right) 
\end{eqnarray}
\par
As we have seen, 
nonrelativistic calculations, based on purely two-body interactions, fail 
to reproduce the correct saturation point of symmetric nuclear matter,
and three-body forces among nucleons are needed to correct this
deficiency. In the sequel the so-called Urbana model will be used, 
which consists, as we have already seen, of an attractive term due to 
two-pion exchange
with excitation of an intermediate $\Delta$ resonance, and a repulsive 
phenomenological central term. 
We introduced the same Urbana three-nucleon
model within the BHF approach (for more details see Ref.~\cite{bbb}).
In our approach the TBF is reduced to a density dependent two-body force by
averaging on the position of the third particle, assuming that the
probability of having two particles at a given distance is reduced 
according to the two-body correlation function. 
The corresponding nucleon matter EOS (no hyperon) satisfies several 
requirements, namely
(i) it reproduces correctly the nuclear matter saturation point,
(ii) the incompressibility  is compatible
with values extracted from phenomenology, 
(iii) the symmetry energy is compatible with nuclear phenomenology, 
(iv) the causality condition is always fulfilled.  
\par
If leptons, namely electrons and muons, and hyperon are introduced,
the general EOS can be calculated for a given composition
of the baryon components.
This allows the determination of the chemical potentials (by simple
numerical derivatives of the energy)
of all the species, baryonic and leptonic, which are 
the fundamental input for the equations
of chemical equilibrium. The latter determines the actual
detailed composition of the dense matter and
 therefore the  EOS to be used in the interior of neutron stars.
Indeed, at high density the matter composition is constrained by 
three conditions: 
i) chemical equilibrium among the different species,
ii) charge neutrality, and iii) baryon number conservation.
At density $\rho \approx \rho_0$ the
stellar matter is composed of a mixture of neutrons, protons,
electrons, and muons in $\beta$-equilibrium [electrons are ultrarelativistic 
at these densities, $\mu_e = (3 \pi^2 \rho x_e)^{1/3}$]. 
In that case the equations for chemical equilibrium read
\begin{eqnarray}
  \mu_n &=& \mu_p + \mu_e \:,
\\
  \mu_e &=& \mu_\mu \:.
\end{eqnarray}
Since we are looking at neutron stars after neutrinos have escaped,
we set the neutrino chemical potential equal to zero. 
Strange baryons appear at density $\rho \approx (2-3) \rho_0$ \cite{bbs}, 
mainly in baryonic
processes like $n + n \rightarrow p + \Sigma^{-}$ and $n + n \rightarrow n + 
\Lambda$.
The equilibrium conditions for those processes read
\begin{eqnarray}
  2\mu_n &=& \mu_p + \mu_\Sigma \:,
\\
  \mu_n &=& \mu_\Lambda \:.
\end{eqnarray}
The other two conditions of charge neutrality and baryon number conservation
allow the unique solution of a closed system of equations, 
yielding the equilibrium
fractions of the baryon and lepton species for each fixed baryon density.
They read 
\begin{eqnarray}
   \rho_p  &=& \rho_e + \rho_\mu + \rho_\Sigma \:,
\label{e:charge}
\\
   \rho &=& \rho_n + \rho_p + \rho_\Sigma + \rho_\Lambda \:.
\label{e:baryon}
\end{eqnarray}
Finally, from the knowledge of the equilibrium composition one determines
the equation of state, i.e., the relation between 
pressure P and baryon density $\rho$. 
It can be easily obtained from the thermodynamical relation 
\begin{equation}
  P = -\frac{dE}{dV}  \:. 
\label{e:press}
\end{equation}
being E the total energy and V the total volume.
Equation (\ref{e:press}) can be explicitly worked out in terms of the 
baryonic and leptonic binding energies, respectively $B$ and $E_L$,
\begin{eqnarray}
  P &=& -\frac{dE}{dV} = -\frac{d}{dV} (B + E_L) = P_B + P_L \:,
\\
  P_B &=& \rho^2 \frac{d(B/A)}{d\rho} = \rho^2 \frac{d}{d\rho}
\left[(x_n + x_p) \frac{\epsilon_{NN}}{\rho_N} + 
x_\Sigma \frac{\epsilon_{N\Sigma}}{\rho_\Sigma} 
+ x_\Lambda \frac{\epsilon_{N\Lambda}}{\rho_\Lambda} \right] \:,
\\  
  P_L &=& \rho^2 \frac{d(E_L/A)}{d\rho} = \rho^2 \frac{d}{d\rho}
\left[x_{e^-} \frac{\epsilon_{e^-}}{\rho_{e^-}} +  x_{\mu^-} 
\frac{\epsilon_{\mu^-}}{\rho_{\mu^-}} \right] \:.  
\label{e:pres}
\end{eqnarray}
In the above equations $x_i$ represent the baryon fraction of 
each species. As far as the leptons are concerned, at those high
densities electrons are a free ultrarelativistic gas,
whereas muons are relativistic. Therefore their energy densities $\epsilon_L$
are well-known from textbooks, see {\it e.g.} ref.\cite{shapiro}.
In order to construct models of neutron stars, one needs to calculate
the total mass-energy density $\cal E$ as well. This can be easily 
obtained just adding the mass-energy densities of each species ${\cal E}_i$
\begin{eqnarray}
{\cal E} &=& {\cal{E}}_{N} + {\cal{E}}_{\Sigma} +  {\cal{E}}_{\Lambda} + 
{\cal{E}}_{e^-} + {\cal{E}}_{\mu^-} \:,  
\end{eqnarray}
While the electron and muon contributions, respectively 
${\cal{E}}_{e^-}$ and ${\cal{E}}_{\mu^-}$, are known from textbooks, 
the baryonic contribution are given by
\begin{eqnarray}
{{\cal E}}_N &=& \frac{1}{c^2}~ (\epsilon_{NN} + m_N \rho_N) \:,
\\ 
{{\cal E}}_\Sigma &=& \frac{1}{c^2}~ (\epsilon_{N\Sigma} + m_\Sigma 
\rho_\Sigma)\:,
\\
{{\cal E}}_\Lambda &=& \frac{1}{c^2}~ (\epsilon_{N\Lambda} + 
m_\Lambda \rho_\Lambda)\:. 
\end{eqnarray}
being $m_i$ the rest mass and $c$ the speed of light. 
For more details, the reader is referred to ref. \cite{bbs} and 
references therein. 
\par
\begin{figure} [ht]
\vspace{2.5 cm}
 \begin{center}
\includegraphics[bb= 90 130 515 694,angle=0,scale=0.6]{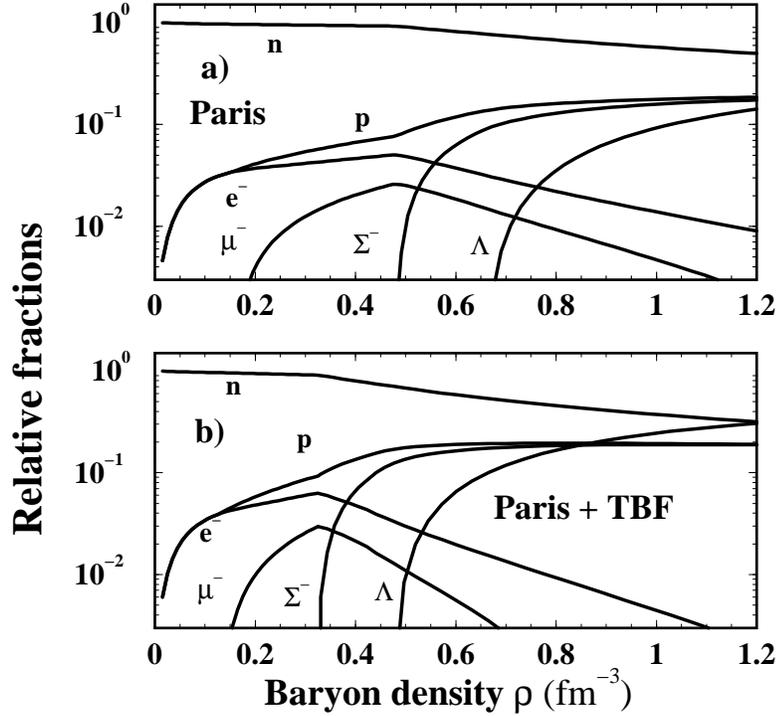}
\end{center}
\vspace{-3.5 cm}
\caption[]{The equilibrium composition of asymmetric and 
$\beta$-stable nuclear matter containing $\Sigma^-$ and $\Lambda$ hyperons
is displayed. In the upper panel only two-body nucleonic forces are present,
whereas in the lower panel TBF's have been included.
\label{cortona1}}
\end{figure}
\begin{figure} [ht]
 \begin{center}
\includegraphics[bb= 90 130 515 694,angle=270,scale=0.5]{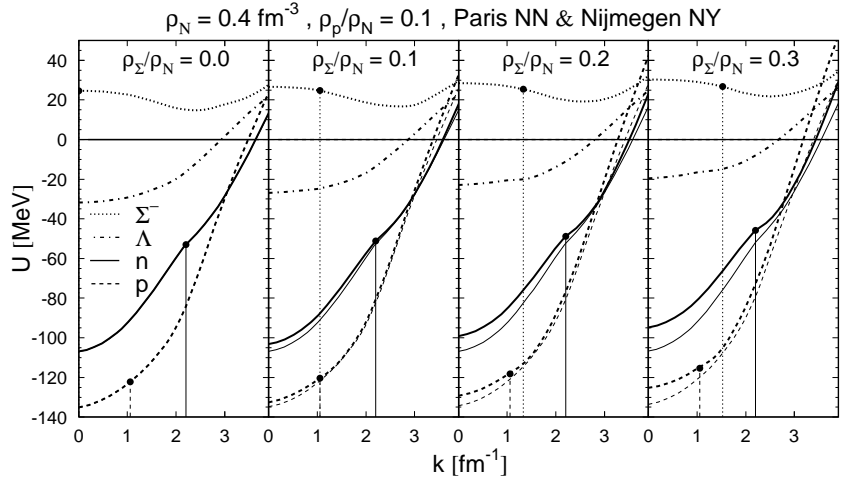}
\end{center}
\caption[]{The single-particle potentials of nucleons $n$, $p$ 
  and hyperons $\Sigma$, $\Lambda$ in baryonic matter of fixed nucleonic
  density $\rho_N=0.4\,\rm fm^{-3}$, proton density $\rho_p/\rho_N=0.1$,
  and varying $\Sigma$ density $\rho_\Sigma/\rho_N=0.0,0.1,0.2,0.3$.
  The vertical lines represent the corresponding Fermi momenta of 
  $n$, $p$, and $\Sigma$. 
  For the nucleonic curves, the thick lines represent the complete
  single-particle potentials $U_N$, whereas the thin lines show the values 
  excluding the $\Sigma$ contribution, i.e., $U_N^{(n)} + U_N^{(p)}$.}
\label{pot}
\end{figure}

In figure~\ref{cortona1} we show the chemical composition of $\beta$-stable 
and asymmetric nuclear matter containing hyperons.
In the upper panel we display the case when only two-body nucleonic 
forces are present, whereas in panel b) nucleonic TBF's are included.
We observe that the inclusion of TBF's shifts the hyperon onset points
down to  $\rho \simeq 2-3$ times normal nuclear matter
density, since some additional repulsion is now present.  
Moreover, an almost equal percentage of nucleons and hyperons are 
present in the stellar core at high densities. A strong
deleptonization of matter takes place, since it is energetically convenient 
to maintain charge neutrality through hyperon formation than $\beta$-decay.
This can have far reaching consequences for the onset of kaon condensation.
The main physical features of the nuclear EOS which determine the
resulting compositions are essentially the symmetry energy of the nucleon 
part of the EOS and the hyperon single particle potentials inside 
nuclear matter. Since at low enough density the nucleon matter is quite 
asymmetric, the small percentage of protons feel a deep single particle
potential, and therefore it is energetically convenient to create 
a $\Sigma^{-}$ hyperon since then a neutron must be converted into a proton.
The deepness of the proton potential is mainly determined by the
nuclear matter symmetry energy. Furthermore, the potential felt by the
hyperons can shift substantially the threshold density at which each
hyperon set in. This points are illustrated in Fig.~\ref{pot},
where the different single particle potentials are plotted at a given
nucleon density.
For simplicity, neutron and proton densities are fixed,
given by $\rho_N=0.4\,\rm fm^{-3}$ and $\rho_p/\rho_N=0.1$,
and the  $\Sigma^{-}$ density is varied.
Under these conditions the  
$\Sigma^{-}$ single-particle potential is sizably repulsive, while
$U_\Lambda$ is still attractive (see also Ref.~\cite{bbs}) and the nucleons
are much strongly bound.
The $\Sigma^{-}$ single-particle potential has a particular shape with an 
effective mass $m^*/m$ slightly larger than 1,
whereas the lambda effective mass is typically about 0.8 and the
nucleon effective masses are much smaller.
\par
\begin{figure} [ht]
\vspace{0.5 cm}
 \begin{center}
\includegraphics[bb= 90 130 515 694,angle=0,scale=0.6]{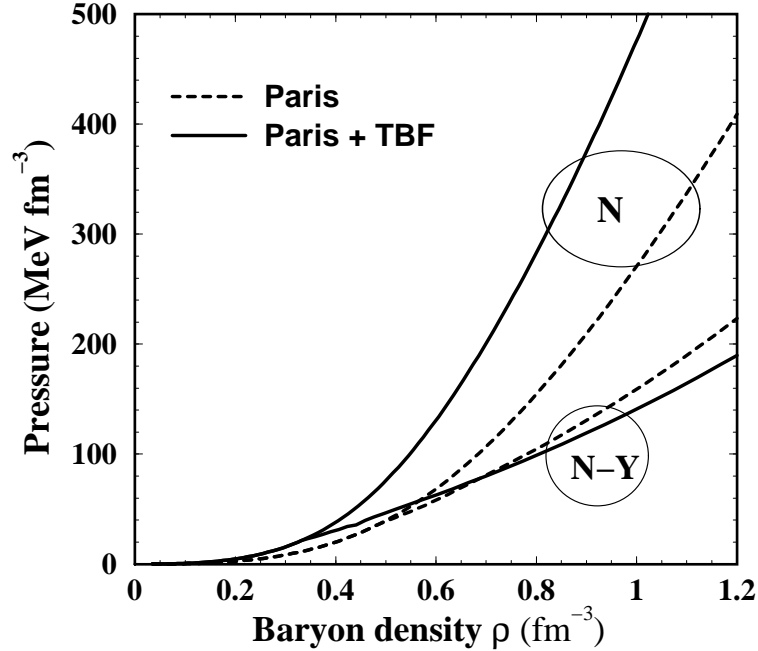}
\end{center}
\vspace{-3.1 cm}
\caption[]{The pressure is displayed vs. the baryon density 
for hyperon-free (upper curves) and hyperon-rich (lower curves) matter.
The solid (dashed) lines represent the case when nucleonic TBF's 
are (are not) included.}
\label{cortona2}
\end{figure}

The resulting Equation of State is displayed in Figure~\ref{cortona2}. 
The dotted
line represents the case when only two-body forces are present, 
whereas the solid line shows the case when TBF's are included.
The upper curves show the equation of state when stellar matter is composed 
only by nucleons and leptons. We mainly observe a stiffening of the equation 
of state because of the repulsive contribution coming from the TBF's.
The inclusion of hyperons (lower curves) produces a soft equation of state
which turns out to be very similar to the one obtained without TBF's.    
This is quite astonishing because, in the pure nucleon case, the repulsive
character of TBF at high density increases the stiffness of the EOS,
thus changing dramatically the equation of state. 
However, when hyperons are included, the presence of TBF's among nucleons 
enhances the population of $\Sigma^-$ and $\Lambda$ because of the increased 
nucleon chemical potentials with respect to the
case without TBF, thus decreasing the nucleon population.
The net result is that the equation of state looks very similar to the case
without TBF, but the chemical composition of 
matter containing hyperons is very different when TBF are included.
In the latter case, the hyperon populations are larger than in the case 
with only two-body forces. 
This has very important consequences for the structure of the neutron stars. 
Of course, this scenario could partly change if hyperon-hyperon 
interactions were known or if TBF would be included also for hyperons, 
but this is beyond our current knowledge of the strong interaction.   

\section{Equilibrium configurations of neutron stars}

\begin{figure} [ht]
\vspace{-1.2 cm}
 \begin{center}
\includegraphics[bb= 90 130 515 694,angle=270,scale=0.56]{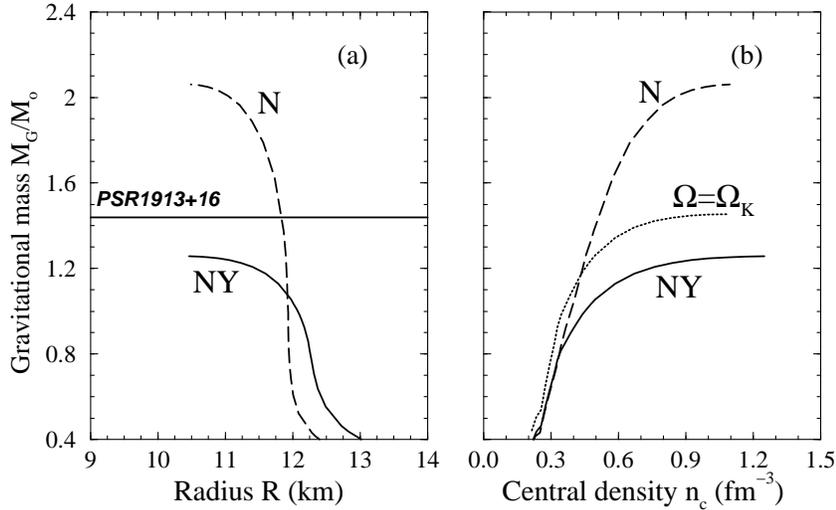}
\end{center}
\vspace{-0.2 cm}
\caption[]{In panel (a) the mass-radius relation is shown in the case of 
beta-stable matter with hyperons (solid line) and without hyperons 
(dashed line). The thick line represents the measured value of the pulsar 
PSR1913+16 mass. In panel (b) the mass is displayed
vs. the central density. The dotted line represents the equilibrium 
configurations of neutron stars containing nucleons plus hyperons and 
rotating at the Kepler frequency $\Omega_K$.}
\label{bolo2}
\end{figure}

We assume that a star is a spherically symmetric distribution of mass in
hydrostatic equilibrium. The equilibrium configurations are obtained
by solving the Tolman-Oppenheimer-Volkoff (TOV) equations \cite{shapiro} for 
the pressure $P$ and the enclosed mass $m$,
\begin{eqnarray}
  {dP(r)\over{dr}} &=& -{ G m(r) {\cal E}(r) \over r^2 } \,
  {  \left[ 1 + {P(r) / {\cal E}(r)} \right] 
  \left[ 1 + {4\pi r^3 P(r) / m(r)} \right] 
  \over
  1 - {2G m(r)/ r} } \:,
\\
  {dm(r) \over dr} &=& 4 \pi r^2 {\cal E}(r) \:,
\end{eqnarray}
being $G$ the gravitational constant. 
Starting with a central mass density ${\cal E}(r=0) \equiv {\cal E}_c$,  
we integrate out until the pressure on the surface equals the one 
corresponding to the density of iron.
This gives the stellar radius $R$ and the gravitational mass is then 
\begin{equation}
M_G~ \equiv ~ m(R)  = 4\pi \int_0^Rdr~ r^2 {\cal E}(r) \:. 
\end{equation}
For the outer part of the neutron star we have used the equations of state
by Feynman-Metropolis-Teller \cite{fey} and Baym-Pethick-Sutherland 
\cite{baym}, and for the medium-density regime  
we use the results of Negele and Vautherin \cite{nv}. 
For density $\rho > 0.08\,{\rm fm}^{-3}$ 
we use the microscopic equations of state obtained in the BHF approximation
described above. For comparison, we also perform calculations of neutron 
star structure for the case of asymmetric and $\beta$-stable nucleonic matter.
The results are plotted in Fig.~\ref{bolo2}.
 We display the gravitational mass $M_G$ 
(in units of the solar mass $M_o$)
as a function of the radius $R$ (panel (a)) and central baryon density 
$n_c$ (panel (b)). We note that the inclusion of hyperons lowers the 
value of the maximum mass from about 2.1 $M_o$ down to 1.26 $M_o$.
This value lies below the value of the best 
observed pulsar mass, PSR1916+13, which amounts to 1.44 solar masses.
However the observational data can be fitted if rotations are included, see
dotted line in panel (b). In this case only equilibrium configurations 
rotating at the Kepler frequency $\Omega_K$ are shown. However,
$\Omega_K$  is much larger than the rotational frequency of that
pulsar, and therefore rotation probably does not play any role. 
\par\noindent
In conclusion, the main finding of our work is the surprisingly low value
of the maximum mass of a neutron star, which hardly comprises the 
observational data. This fact indicates how sensitive the properties of the
neutron stars are to the details of the interaction. In particular
our result calls for the need of including realistic hyperon-hyperon 
interactions. However, the use of the available hyperon-hyperon interactions
seem to introduce only minor changes in the results  \cite{barc}.
Despite the uncertainty on the NY and YY interactions,
it is unlikely that one can obtain a neutron star mass substantially 
larger. The possible occurrence of a quark core is usually assumed to
further soften the EOS and lower the maximum mass. However,
this is not necessarily true, since at large density the quark pressure
should be close to the one of a ultra-relativistic Fermi gas, which
can rise fast enough to stabilize the system. In any case, the possible
quark core is not expected to change dramatically the critical neutron 
star mass. Even if an explicit analysis of the quark core has still
to be worked out, it is fair to say that
the observation of a neutron star with a mass much larger than
1.4-1.5 solar mass would indicate that indeed some basic ingredient is missing
in our understanding of neutron star structure.  

\section*{Acknowledgments}

The material presented in this contribution is the result of a
fruitful collaboration, lasting for several years, with a number of people.
Special thanks are due to Dr. I. Bombaci, Prof. L.S. Ferreira, 
Dr. G. Giansiracusa, Prof. U. Lombardo, Dr. H.-J. Schulze and Prof. H.Q. Song.


\begin{thebibliography}{8.}
\addcontentsline{toc}{section}{References}

\bibitem{book} For a pedagogical introduction, see 
 {\it Nuclear Methods and the Nuclear Equation of State}, 
 Edited by M. Baldo, World Scientific, Singapore, International Review of
 Nuclear Physics Vol. 9, 1999.

\bibitem{Day} B.D. Day, {\it Brueckner--Bethe Calculations of Nuclear Matter},
Proceedings of the School E. Fermi, Varenna 1981, Course LXXIX,
ed. A. Molinari, (Editrice Compositori, Bologna, 1983), p. 1--72;
{\em Rev. Mod. Phys.} {\bf 39}, 719 (1967).

\bibitem{bbp} H.A. Bethe, B.H. Brandow and A.G. Petschek,  Phys. Rev.
  {\bf 129}, 225 (1962) 

\bibitem{mahaux} J. H\"ufner and C. Mahaux, {\it Ann. Phys. (N. Y.)}
{\bf 73}, 525 (1972).

\bibitem{v14} R.B. Wiringa, R.A. Smith and T.L. Ainsworth,
   {\em Phys. Rev.} C{\bf 29}, 1207 (1984).

\bibitem{Fadeev} L.D. Fadeev, {\it Mathematical Aspects of the 
Three-Body Problem in Quantum Scattering Theory}, Davey, New York 1965.

\bibitem{Bethe} H.A. Bethe, Phys. Rev. {\bf 138}, 804 (1965).

\bibitem{Raja} R.Rajaraman and H.Bethe, {\em Rev. Mod. Phys.} 
{\bf 39}, 745 (1967).

\bibitem{Day1980} B.D. Day, {\em Phys. Rev.} C{\bf 24}, 1203 (1981);
{\em Phys. Rev. Lett.} {\bf 47}, 226 (1981);

\bibitem{song} H.Q. Song, M. Baldo, G. Giansiracusa 
and U. Lombardo, {\em Phys. Rev. Lett.} {\bf 81}, 1584 (1998).

\bibitem{v18} R.B. Wiringa, V.G.J. Stocks and R. Schiavilla, Phys. Rev. 
 C{\bf 51}, 38 (1995).

\bibitem{Bonn} R. Machleidt, {\it Adv. Nucl. Phys.} {\bf 19}, 189 (1989).

\bibitem{earl3bf}J. Fuyita and  H. Miyazawa, Progr. in Theor. Phys., 
 {\bf 17}, 360 (1957).

\bibitem{Sauer} Ch. Hadjuk, P.U. Sauer and W. Streuve, Nucl. Phys.
 {\bf A405}, 581 (1983).

\bibitem{Gran}P. Grang\'e, A. Lejeune, M. Martzolff and J.-F. Mathiot, 
 Phys. Rev {\bf C40}, 1040 (1989), and references therein. 

\bibitem{Math} J.-F. Mathiot, Phys. Rep. {\bf 173}, 63 (1989).

\bibitem{BrownWeise} G.E. Brown, W. Weise, G. Baym and J. Speth,  
      {\it Comm.\ Nucl. Part. Phys.} {\bf 17}, 39 (1987).
\bibitem{3bfUrb} Carlson J., Pandharipande V.R. and Wiringa R.B., 
 Nucl. Phys. {\bf A401}, 59 (1983).

\bibitem{neumat} M. Baldo, G. Giansiracusa, U. Lombardo and H. Q. Song,
    {\em Phys. Lett.} B{\bf 473} (2000) 1.

\bibitem{umb1} W. Zuo, G.Giansiracusa, U. Lombardo, N. Sandulesco and
 H.-J. Schulze, Phys. Lett. {\bf B421}, 1 (1998).  

\bibitem{umb2} W. Zuo, U. Lombardo, and H.-J. Schulze, 
 Phys. Lett. {\bf B432}, 241 (1998).  

\bibitem{mae89}
   P. Maessen, Th. Rijken, and J. de Swart,
   {\em Phys. Rev.} C{\bf 40}, 2226 (1989).

\bibitem{bbb}
   M. Baldo, I. Bombaci, and G. F. Burgio,
   \Journal {\A&A}{328}{274}{1997}.

\bibitem{bbs}
   M. Baldo, G. F. Burgio, and H.-J. Schulze,
   \Journal{\PRC}{61}{055801-1}{2000}. 

\bibitem{shapiro}
   S. L. Shapiro and S. A. Teukolsky, 
   {\em Black Holes, White Dwarfs and Neutron Stars}  
   (John Wiley \& Sons, New York, 1983).

\bibitem{fey}
   R. Feynman, F. Metropolis, and E. Teller,
   \Journal{\PRC}{75}{1561}{1949};

\bibitem{baym}
   G. Baym, C. Pethick, and D. Sutherland,
   \Journal{\ApJ}{170}{299} {1971}.

\bibitem{nv}
   J. W. Negele and D. Vautherin,
   \Journal{\NPA}{207}{298}{1973}.

\bibitem{barc}
   I. Vida\~na, A. Polls, A. Ramos, L. Engvik, and M. Hjorth-Jensen,
   preprint University of Barcellona, 1999.

\end{thebibliography}
\end{document}